\documentclass[epj]{webofc}
\usepackage[varg]{txfonts} 
\wocname{EPJ Web of Conferences}
\woctitle{ICNFP 2016}
\newcommand {\bi} {\bibitem}
\newcommand {\be} {\begin{equation}}
\newcommand {\ee} {\end{equation}}
\newcommand {\bea} {\begin{eqnarray} }
\newcommand {\eea} {\nonumber \end{eqnarray}}
\newcommand {\eps} {\epsilon}

\newcommand {\bc} {\begin{center}}
\newcommand {\ec} {\end{center}}
\newcommand {\bd}{\begin{displaymath}}
\newcommand {\ed}{\end{displaymath}}

\def \form#1 {eq. (\ref{#1}) }
\def \parziale#1#2 {{\partial {#1} \over \partial {#2}}}
\def \bi#1 {\typeout{#1} \item}

\begin{document}
\selectlanguage{english}
\title{Historical and personal recollections of Guido Altarelli}
%
%

\author{Giorgio Parisi\inst{1,2,3}\fnsep\thanks{\email{Giorgio.Parisi@Roma1.Infn.It} } 
}

\institute{Dipartimento di Fisica, {\it Universit\`a degli studi di Roma La Sapienza}, 
\and
 Nanotec-CNR, UOS Rome,
\and
 INFN-Sezione di Roma 1, Piazzale A. Moro 2, 00185, Rome
}

\abstract{%
 In this paper I will present a short scientific biography of Guido Altarelli, briefly describing some of his most important seminal works. I will analyze in great details the paper of the $q^2$ evolution of the effective quark distribution: I will put this paper in a historical perspective, describing our theoretical understanding at that time and the reasons why the paper was so successful.
}
\maketitle

\section{A short scientific biography}

Guido Altarelli, a great theoretical physicists, who spent most of its scientific career in Rome and at CERN, was born in Rome on 12 July 1941 and passed away on 30 September 2015 in his 74th year. 

He received three very prestigious awards: {the Julius Wess Award (2011), the J. J. Sakurai Prize of the APS (2012) and the High Energy and Particle Physics Prize of the EPS (2015)}. These three very celebrated distinctions only partially reflect the richness of his contributions: his deep and lasting influence is ubiquitous in the field of high energy physics.

 Guido graduated in Physics from {Rome University} in 1963, the supervisor being Raul Gatto (who was in Florence at that time). 
 In 1964 he went to {Florence} to join the very lively and large group of young researchers ({\sl gattini}, i.e. small cats) who were working together under the careful and inspiring supervision of Gatto\footnote {In Italian "gatto" means "cat".}. 
 In 1968-1970 he spent two years in {New York} (at New York University and at Rockefeller University).
 In 1970 he became professor of theoretical physics in {Rome}. In this period he spent long periods in other institutions, e.g. the ENS in Paris and  the Boston University. 
 
 In 1987 Guido became Senior Staff Physicist at the {Theory Division of CERN} (where remained up to the age of 65 years in 2006), and he was the Theory Division Leader in 2000-04. In 1992 he moved to the newly founded University Rome 3, where he remained up to formal retirement, dividing his time between teaching (in Rome) and research (mostly at CERN). He continued his research activity up to his last months. 

 As usually in Italy, his first research work was his {Laurea} thesis: it was done in collaboration with his friend Franco Buccella on the process $e^++ e^- \to e^++ e^- +\gamma$. Gatto suggested them to perform this computation in preparation for future experiments at the ADONE colliding beams of Frascati; it was part of a wide spectrum program lead by Nicola Cabibbo and Gatto, aimed to compute the cross sections for every possible process that could be observed \cite{CabibboGatto} in  $e^++ e^-$ colliding beam experiments. Guido Altarelli and Franco Buccella had to evaluate the cross section for the process  $e^++ e^- \to e^++ e^- +\gamma$, paying attention to the one-loop infrared divergences in the process $e^++ e^- \to e^+ +e^-$. The formulae they wrote in their first attempt were correct. Unfortunately, for very small transfer momentum of the $\gamma$ the computer was nearly computing 0/0 and, due to rounding effects in floating point arithmetics, the output was extraordinarily large. This first attempt produced an extremely large cross section that left everybody puzzled: however they were fast enough to understand the reason of the strange result: they finally wrote a successful thesis \cite{ABuccella}.
 
 In Florence he worked with Gatto and the other gattini (among them Marco Ademollo, Buccella, Luciano Maiani and Giuliano Preparata). Most of the topics they investigated were fashionable at that time: e.g. phenomenological applications (mainly to to hadron spectroscopy) of various symmetries ($SU(6)$, $U(6) \otimes U(6)$ and $U(12)$ )\cite{ABC,AAG,AGMP}; he also started to be interested in Regge poles \cite{REGGE}. He continued to work on the theoretical evaluations of the cross section of different reactions in electron-positron colliding beams.
 
 In New York, he worked mainly with Hector Rubinstein on Regge pole predictions coming from the resummation of ladder diagrams, on duality and on bootstrap \cite{AR}. At that time the dominant school was asserting that the mystery of strong interaction could be understood using only general self-consistency properties of the $S$ matrix ($S$ matrix bootstrap). A quantum field theory approach was considered pointless: I remember that in some circles a too deep knowledge of quantum field theory was considered to be a handicap. This phenomenological program culminated in the duality and in the superconvergence sum rules. This line of research reached a definitive success with the Veneziano model \cite{VENEZIANO} (and a few month later with the Virasoro model \cite{VIRASORO}). It is amusing to note that these two models were the starting point of the most sophisticated existing field theory, i.e. string theory; they correspond respectively to the open and to the closed string models.
 
After his arrival in Rome, his scientific interests moved firstly toward the parton model and later concentrated on what we now call the Standard Theory. This change was in agreement with the strong tradition of the Rome school (led by Cabibbo who became a full professor at the Rome University in 1966 at the age of 31), of using field theory tools to understand particle physics: it was the alternative viewpoint respect to the $S$ matrix bootstrap. 

Cabibbo was deeply interested in weak interaction physics. The theory of weak interaction is based on local currents, Lagrangians: one needs the whole the apparatus of local quantum field theory and Cabibbo's interest in field theory was a necessity. He created a strong school of younger collaborators and students and the Rome school gave a crucial contribution to the construction of the standard model.

Many of the Roman works of Guido were written with Cabibbo, Mario Greco, Keith Ellis, Maiani, Guido Martinelli, Giorgio Parisi and Roberto Petronzio. 
Among the most important contributions are the papers on the octet enhancement of non-leptonic weak interactions in asymptotically free gauge theories \cite{OCTECT,OCTECT1}: for the first time it was shown that the newly born QCD was able to contribute to solving some of the old mysteries of weak interactions (in this case the dominance of the $\Delta I=1/2$ strangeness changing processes over the much weaker $\Delta I=3/2$ processes). 

Other important contributions of this period are:
\begin{itemize}
\item His paper \cite{AP} on the equations for the $q^2$ dependence of the proton structure functions where effective parton model distributions were introduced and the equation for the $q^2$ evolutions were written (the so called DGLAP equations).
\item The first computation of the next to the leading order (NLO) corrections to the na\"ive parton model \cite{AEM}: in this paper he discovered that there are quite large one-loop NLO corrections to the standard parton prediction for the $\mu^++\mu^-$ production in hadronic corrections: there is a correction factor of order 1 at not extremely large values of the invariant mass of the muon pairs. 
\item The first computation of the transverse momentum distribution of the $\mu^++\mu^-$ pairs produced in hadronic collisions \cite{transverse}. This was an important check of QCD: one could see directly the effects of hard scattering of the partons that producing very large momentum pairs. The process also provided an independent way to measure the running strong coupling constant.
\item A careful analysis of the production of vector mesons at colliders where theoretical QCD based considerations mix with more phenomenological considerations \cite{vector}. This paper was crucial to construct a well-founded model for the production of vector mesons as a function of total energy, transverse momentum, and rapidity.
\end{itemize}
 
 In the 70's Guido and I have been working side by side: also when I was not actually working together with him I was quite aware of his works. This went on up to the beginning of the 80's. At that time I became mostly interested in problems in statistical mechanics and my only residual activity in high energy physics was lattice QCD. Guido, especially after going to CERN, started to explore new territories in high energy physics and our roads unfortunately separated.
 
 His interaction with CERN's environment strongly reflected on his choice of research subjects. He gave crucial contributions to the understanding of the physics of the colliders: the planning of experiments in CERN was strongly influenced by his theoretical analysis. His work has also been crucial in putting an order into the huge mass of experimental data and theories (sometimes one contradictory with the other) and in identifying the crucial experiments or theoretical analysis that had to be done to remove the contradictions.
 
The number results that he obtained in this period is so abundant that I will only make an arbitrary selection,  strongly biased by my personal tastes:
\begin{itemize} 
\item In the 80's he got interested in predictions {for future CERN experiments}, e.g. for the production of jets, heavy-vector mesons \cite{ADMN} and other exotic objects like supersymmetric particles \cite{SUSY} and the Higgs meson \cite{H2}. Of course, the production of the Higgs meson is strongly dependent on his mass, so it was crucial to get the most accurate bounds on his mass from the comparison of the experimental data with the standard model \cite{H1}.
\item He has also written seminal papers on the decay of heavy quarks (e.g on inclusive beauty decays and on the spectator model \cite{spectator}). In this context he continued the analysis of the {consequences of QCD on weak interactions}, computing the two loops contributions \cite{Altarelli:1980fi}: it was the first NLO computation for the strong corrections to weak interactions.
\item He became to be deeply interested in polarized proton structure function \cite{POL}, where he discovered, together with Graham Ross \cite{ROSS}, {the crucial interplay between the gluon anomaly and polarization effects.}
\item Mainly with Riccardo Barbieri, he constructed a model-independent analysis of the electroweak data \cite{B1, B2, B3, B4}. This approach was quite new and the methodology of this study became a standard for future analysis.
\end{itemize}
 
 With the new millennium, Guido starts to study new subjects. He is mostly interested in the attempt of deciphering the information contained in the huge mass of experimental data, especially in the newly obtained mass matrices of neutrino. He looks for fingerprints of symmetries; his old expertise in symmetry groups (from Florence times) has certainly been very useful to him.
 \begin{itemize}
 \item He studies the detailed predictions of realistic $\mbox{SU}(5)$ unification (also with supersymmetry) both for proton decay \cite{SU5} and for neutrino mixing \cite{SU5M}.
 \item He is fascinated by the elegance of the {tribimaximal neutrino mixing}: many of his papers \cite{F} (mostly with Ferruccio Feruglio) are dedicated to the search for the origins of this baffling symmetry. Although nowadays we know that neutrino mass matrix is not tribimaximal, it is still possible that deviations from a tribimaximalily come from a perturbative expansion in some small parameter.
 \item He is still interested to the evolutions of parton distribution, especially in the small $x$ region, an important experimental region, where many data are available. The theoretical analysis is quite complex, however, his careful analysis shows that the resummation of the leading logarithms in $x$ gives a precise prediction that is in very good agreement with the experiments \cite{X}. 
\end{itemize}

\section{The effective $q^2$ dependent parton distribution functions and their evolution equations}
 Here I will present some historical recollections on the paper {\it Asymptotic freedom in parton language} that we wrote when Guido and myself were both in Paris (Guido was in the ENS in the center of Paris and I were in the IHES at Bures-sur-Yvette). I will concentrate my attention on this paper also because of it the paper of Guido I know better; I believe that I am qualified to describe the motivations and the perspectives. I will not discuss other people's important works \cite{GL, D} in the same direction that we were not aware of at that time.

 \subsection{Field theories preliminaries}

Generally speaking, the relevance of a paper can be understood only in the context of the physics that was known at the time that the paper is written. One should be very carefully: if we miss the historical context when reading a paper after a long time, we risk to consider trivial the main advances of that paper and to consider very relevant something that we quite trivial. The historical reconstruction of the knowledge and of the understanding at the time when the paper was written is crucial to avoid misinterpretations.

 At the time of the paper, very few things were known in strongly interacting quantum field.
 About 10 years before Murray Gell-Mann wrote \cite{GM} in a very influential paper: {\sl We use the method of abstraction from a Lagrangian field theory
model. In other words, we construct a mathematical theory of
the strongly interacting particles, which may or may not have
anything to do with reality, find suitable algebraic relations that
hold in the model, postulate their validity and then throw away
the model. We compare this process to a method sometimes
employed in French cuisine: {\it a piece of pheasant meat is cooked
between two slices of veal, which are then discarded}}.

 As remarked by David Gross {\sl until 1973 it was not thought proper to use field theories without apologies} \cite{GROSS}. Feynman himself refrained to speak of field theory and invented  partons. In 1977 the situation was much better; however, people were cautious with field theory: they were very careful in making assumptions that were not proved at least at all orders in perturbation theory and nearly nothing was proved.
 
 One of the most solid results was Wilson \cite{WILSON} short distance operator product expansion that states that in the limit of small $x$ the product of two operators can be written as
\be
A(x) B(0) \to_{x\to 0} \sum_C C(0) |x|^{-d_A-d_B+d_C} \,,
\ee
where the sum runs over all the local operators $C$, including the identity. The leading terms come from the operators $C$ with the lowest dimensions.
The dimensions of the operators were the canonical one in free field theory and this simple case everything was clear. 

Strong interactions are not a free field theory. In the case of a non-free theory, we have to use the renormalization group. 
The dimensions of the operators could be different from the canonical ones in an interacting theory. In a seminal paper \cite{S} Symanzik studied the case of a $\phi^4$ theory: he obtained
\be
\phi(x) \phi(0) \to_{x\to 0} \alpha |x|^{-2d_\phi}{\bf 1}+\beta |x|^{-2d_\phi+d_{\phi^2}} \phi^2(0)\,,
\ee
where ${\bf 1}$ is the identity operator.

In this case, the Wilson expansion was proved, and the dimensions were not the naive one, provided that the renormalization group had a non-trivial fixed point: the dimensions contained a non-trivial coupling constant dependent part, the called anomalous dimensions. The renormalization group with a running coupling constant has to be used in order to compute these anomalous dimensions.

In order to explain the importance of Symanzik's paper, I have to show its relevance of the short distance operator expansion on deep inelastic scattering. 
 The total cross section for the process "virtual gamma+ proton" (i.e. deep inelastic scattering) can be written in terms of the appropriate Fourier transform of the function:
\be
G(x^2,x_0)=\langle p|J(x)J(0)|p\rangle \,,
\ee
where $|p\rangle$ denotes the one proton state and we have suppressed the Lorentz indices of the currents.
Kinematical considerations imply that the region relevant for deep inelastic scattering is $x^2 \approx 0$, i.e. the light cone \cite{BP}.

At the end of the sixties, it was shown experimentally that in inelastic electron-proton scattering in the region of large momentum transfer, the data were compatible with a simple scaling of the structure functions (the so-called Bjorken scaling). 
 The deep roots of Bjorken scaling for deep inelastic scattering have been soon identified. Bjorken scaling was shown to be a consequence of the naive light cone expansion \cite{BP} of Brandt and Preparata (1969)
\be
J(x)J(0)\to_{x^2\to 0} {O(x_0,0)\over x^2}\,,
\ee
where $O(x_0,0)$ is a bilocal operator. 

This  expansion has the same form of the one valid in a free field theory.
Indeed in free field theories, the naive light cone expansion can be derived by the Wilson expansion by doing a Taylor expansion at $x_0=0$.
 An infinite number of terms in the Taylor expansion have to be considered, each term corresponding to a different moment of the experimental structure function. In this way, the Wilson short distance operator product becomes deeply related to the experimentally observed approximate Bjorken scaling.
 
 However, the crucial question was what happens for an interacting theory? Christ, Hasslacher and Muller (1972) computed the coupling dependent anomalous dimensions of the operators relevant for deep inelastic scattering (the so-called twist-two operators) for a Yang-Mills theory \cite{CM}.

 If we neglect the dependence of the running coupling constant ($\alpha$) on the momenta, when we consider only non-singlet contribution (the gluon contribution is a singlet) they obtained something like
 \begin{equation}\nonumber
M_n(q^2)\equiv\int _0^1dxx^{n-1}F(x,q^2) \ ; \ \ \ M_n(q^2)=C_n \exp ( \gamma_n (\alpha) \log(q^2)) \,.\end{equation}
Here $F(x,q^2)$ is the structure function of the proton, that enters in the computation of deep inelastic scattering.

 If we specialize the Christ-Hasslacher-Muller formulae to QCD, we get a simple result always in the case of the non-singlet contribution. Taking care of the running coupling constant $\alpha(q^2)$, we get the final formulae:
 
 \be {\partial M_n(q^2)\over \partial \log(q^2)}= \gamma_n (\alpha(q^2))\, ,\ee
 where the linear term in $\alpha(q^2)$ of is obtained from formulae of \cite{CM}.
 
If we use the standard properties of the Mellin trasform and of the inverse Mellin we finally get 
\be{\partial F(x,q^2)\over \partial \log(q^2)}=\int_x^1 {dy\over y} F(x,q^2) P_{q,q}(x/y,\alpha(q^2))\,,\ee
where the quark fragmentation function $P_{q,q}(z,\alpha(q^2))$ is the inverse Mellin transform of the anomalous dimensions $\gamma_n (\alpha(q^2))$.
Using the first order in $\alpha$ for $\gamma_n (\alpha)$ computed in \cite{CM}, we finally get for the non singlect contribution:
\be
P_{q,q}(z)=\frac83{\alpha(q^2)\over 4 \pi}\left( {1+z^2\over(1-z)_+} +\frac32 \delta (z-1)\right)\ee

 {When we wrote our work (spring '77), all that was very well known.} In '73 I obtained the previous formula for the valence quarks \cite{GP1}. Later on, similar formulae were written for the gluons and sea-quarks contribution using inverse Mellin transform \cite{PP, MORIOND}.

 \subsection {The parton model}

In theories where the interaction could be neglected in the large momentum transfer region, deep inelastic structure functions measure the number of charged partons (quarks in QCD) carrying a fraction of momentum $x$ in the infinite momentum frame. This observation leads to Bjorken scaling in the parton model. Guido was familiar with the details of the parton model construction and its interpretation in $p=\infty$ frame and he wrote an influential paper on it \cite{NUCLEON}. 

However, the Bjorken scaling was not exact and the effect of the strong interactions in the large momentum region was weak, but it was not vanishing. At that time was a big question how to deal with the parton model concepts in QCD \cite{POLYAKOV, KS}. Many phenomenological arguments were simply formulated using the parton language and the QCD results were obtained via operator product expansion. Most of the relations coming from the parton model where out of reach of the operator product expansion.

 However, there was a different case where the physical interpretation was quite clear: quantum electrodynamics, where electromagnetic radiative corrections produce perturbative effects that are not negligible, i.e. they are not simply proportional to $\alpha$ with a prefactor of order 1. In QED one often finds that the net effect of radiative corrections is proportional to $\alpha \log(E/m_e)$ or $(\alpha \log(E/m_e))^k$. Various techniques can be used.
 The most famous approach that leads to a simple and fast computation of these effects is the {\sl Equivalent photon approximation in quantum electrodynamics} (due to Weizsaker-Williams.).

Nicola Cabibbo was strongly interested in computing these effects for applications to the experimental analysis of events collected in $e^+e^-$ collisions.
Cabibbo and Rocca in 1974 wrote \cite{CABIBBOROCCA} formulae like:

\begin{equation} {P_{e\to e\gamma}(\eta)}=\frac{\alpha}{\pi}{1+(1-\eta)^2\over \eta} \log( E/m_e)\, ; \ \ \ P_{\gamma\to e^+ e^-}(\eps)=\frac{\alpha}{2\pi}(1+(1-2\eps)^2) \log( E/m_e)
\end{equation}
where $\eta$ is the fraction of longitudinal momentum carried by the photon
and $\eps$ is the fraction of longitudinal momentum carried by the electron.

 {These probabilities can be combined.} They computed the probability of finding inside a $\gamma$ a triplet $\gamma, e^+,e^-$: it is proportional to $P_{\gamma\to e^+ e^-} P_{e\to e\gamma}\log( E/m_e)^2$. Nicola and I \footnote{This computation was never published: Cabibbo and Rocca wrote \cite{CABIBBOROCCA} that I did such a computation and I wrote that the computation was done by Cabibbo \cite{MORIOND}. I remember that I had many discussions with Nicola on these generalized Weizsaker-Williams relations, but I forgot what exactly happened in this case. }
 used these ideas \cite{CABIBBOPARISI} to estimate the cross section for the process $ e^+e^- \to 2e^+2e^-$ that was a candidate to be an important background in $e^+e^-$ colliding beam experiments. 

\subsection{The paper}
 This paper stemmed from a proposal from Guido, i.e. to make previously-obtained results on scale violations clearer and more exploitable. It was written while both authors were in Paris, and Guido liked to remark that it is the most cited {\it French} paper in the field of high energy physics. 
 
The motivations of the paper were clearly stated in the introduction. 
 When speaking of the standard operator product approach to scaling violations in QCD, we wrote:
{\sl In spite of the relative simplicity of the final results,
their derivation, although theoretically rigorous, is {somewhat abstract and formal}, being formulated in the language of renormalization group equations for the coefficient functions of the local operators which appear in the light cone expansion for
the product of two currents.}

We also added:
{\sl In this paper, we show that a {alternative derivation} of all results of current interest for the $Q^2$ behavior of deep inelastic structure functions is possible. In this approach {all stages of the calculation refer to parton concepts} and offer a very illuminating physical interpretation of the scaling violations. In our opinion the present approach, { although less general, is remarkably simpler} than the usual one since all relevant results can be derived in a direct way from the basic vertices of QCD, {with no loop calculations} being involved (the only exception is the lowest order expression for the running coupling constant which we do not rederive).
This method can be described as an appropriate generalization of the equivalent photon approximation in quantum electrodynamics (Weizsaker-Williams .... Cabibbo-Rocca).}
 
 {In spring '77} Guido and I were often discussing QCD scaling violations. Guido suggested that it would be pedagogically useful to derive all the equations for scaling violations using the same techniques of Cabibbo-Rocca; { no loops were involved: only the evaluation of the vertices in the infinite momentum frame.} As I remarked before, Guido was an expert on parton model in the infinite momentum frame and the Cabibbo Rocca computation was already done in the infinite momentum. Most of the formulae we needed were written there: only the gluon splitting into two gluons function was missing. However, in our paper, we rederived all the spitting functions.
 
 The final master \footnote{We used the wording {\em master equations} because they were classical probabilistic equations derived in a quantum setting.} equations are well known:
 \begin{eqnarray}
 {dq^i(x,t)\over dt}={\alpha(t)\over 2 \pi}\int _x^1 {dy\over y}\left[q^i(y,t)P_{q,q}\left({x\over y}\right)+G(y,t)P_{q,G}\left({x\over y}\right)\right]\,, \nonumber \\
 {dG^i(x,t)\over dt}={\alpha(t)\over 2 \pi}\int _x^1 {dy\over y}\left[\sum_{i=1}^{2f}q^i(y,t)P_{G,q}\left({x\over y}\right)+G(y,t)P_{G,G}\left({x\over y}\right)\right]\,,
\label{APEQ}
\end{eqnarray}
 where $t=\log(q^2)$ and the index $i$ runs over quarks and antiquarks of all flavour.
 
 We eventually obtained the expression for all the splitting functions ($P_{q,q}, P_{q,G},P_{G,q}, P_{G,G}$). Our computations were much simpler that the original computations based on the one loop corrections to the vertices of the operators entering in the Wilson product expansion of two currents (a l\`a Christ-Hasslacher-Muller). 
The computations were particular transparent and simple when we extended it to the case of polarized partons as an using the helicity formulation in the infinite momentum frame.

The paper was a well-done cocktail of renormalization group results, parton model, and perturbation theory in infinite momentum frame: easy to drink and to swallow. {The paper was very successfully, also because it was very clearly written: it was written by Guido}, not by myself. \ It was really pedagogic: it contained all the logical steps and the final receipt was quite easy to follow.

\subsection{Aftermath}

I am convinced that the most important result of the paper was {\sl not} the construction of a practical way to compute scaling violations in deep inelastic scattering. Indeed this task was already accomplished using the Mellin transformation and related techniques: these techniques were slightly more cumbersome but they were easy to use. Moreover, the equations (\ref{APEQ}) were well known, at least for the valence quarks. 

The important point in our paper was to shift the focus from Wilson operator expansion to the effective number of partons that was dependent on the resolution, i.e. $q^2$. {It was more than a computation: it was a {\sl change in the language} we use. The appropriate choice of the language is one of the most important scientific tools.

For example, the Drell-Yan process (i.e. $pp \to l^+ l^-+\cdots$) could be not studied by a Wilson operator expansion: the Brandt-Preparata analysis did not work in this case. Similar conclusions are also true for the jet production in hadronic collisions and for the computation of the cross section for large transverse momenta that were related to the hard scattering of partons.

However, it was now possible to study these process using the new language: we have to factorize the amplitude for the process into a part containing the effective parton distribution at the relevant energy and into a part containing the hard scattering that could be treated in perturbation theory in the running coupling constant. Moreover, in presence of energy dependent effective parton distributions the parton model predictions were ambiguous at order $\alpha_{QCD}$ and next to the leading order corrections (NLO) have to be computed in order to refine the predictions.

 The solution of all these problems was at hand after our paper. This opportunity was immediately taken by Guido. All the following papers are published in '78 just after our '77 paper:
\begin{itemize}
\item{\it Leptoproduction and Drell-Yan processes beyond the leading approximation in chromodynamics} \cite{AEM}. This is the first NLO computation we can find in the literature: it was simply done by computing the first corrections in $\alpha_{QCD}$ to the process $q+\bar{q}\to l^+ l^-+\cdots$ and comparing them to the first order for deep inelastic scattering. In the experimental region, the corrections were quite surprisingly of order 1, because of a factor $\pi^2$ that was present only at positive $q^2$ and not at negative $q^2$. 

\item{\it Transverse momentum of jets in electroproduction from quantum chromodynamics} \cite{ALMA}. The large transverse momentum distribution of jets in deep inelastic scattering was analytically computed and successfully compared to the experiments. These large transverse momentum events have the same origin of the hard scattering processes that induce scaling violations in deep inelastic scattering.

\item{\it Transverse momentum in Drell-Yan processes} \cite{transverse}. There are two papers on this subjects that are nearly simultaneous to the previously mentioned papers. Here it was shown that the transverse momentum distribution of large mass muon pairs produced in proton-nucleus collisions can be quantitatively explained the combination of two ingredients: the hard scattering of partons (gluon and quarks) and the intrinsic transverse distribution of constituent partons inside the hadrons.

\item{\it Processes involving fragmentation functions beyond the leading order in QCD} \cite{ASO}. Here the whole analysis is based on energy dependent effective parton distributions and in the comparison between the na\"ive parton model and the one loop results.
\end{itemize}

In all these four problems the matching of the experimental data with the theoretical predictions allowed and independent measure of the QCD running coupling constant. At that times it was possible to achieve this goal in the first three cases. The compatibility of these independent determinations of $\alpha_{QCD}$ with the value coming from deep inelastic scattering (and from $\psi$ decay) was instrumental in convincing physicists that QCD was the correct theory.

 \section{We miss him very much}
 {His scientific success was inseparable from his human qualities. } Guido aimed to get a deep understanding of the world, hence his great passion for history, especially that of the many countries he was traveling through.

Perhaps his most characteristic traits were his {great kindness and intellectual honesty, coupled with a rather ironic view of himself and life in general.} His great inquisitiveness, the enjoyment he derived from learning new things and putting the pieces of a puzzle together, allowed him to make great summaries of topical subjects, which allowed us to take stock of the current state of a field of research, {indicating new directions to take. }

 {He liked clear, precise formulations, which could be understood by all.}
He was not a reclusive or selfish scientist, interested only in what personal prestige might be gained from his research. Guido was also a researcher who enjoyed to work with others within the large community of high-energy particle physicists.

 Many of Guido's works, from the most famous to the lesser known, were conceived in a spirit not only of research but also in a spirit of service to the community. {It is difficult to think which would be the status of the field without his {\sl seminal} contributions.}

 We all miss him very much, not only as an invaluable scientist but also as a dear friend, always ready to help.

\end{document}